\documentclass[nonacm, sigconf]{acmart}

\acmBooktitle{}
\acmYear{}
\acmMonth{}
\acmConference{}{}{}

\renewcommand\footnotetextcopyrightpermission[1]{}
\acmPrice{}
\acmISBN{}
\acmDOI{}

\AtBeginDocument{%
  }

\usepackage{float}
\usepackage{hyperref}

\begin{document}

\title{\textit{Can You Keep a Secret?} Exploring AI for Care Coordination in Cognitive Decline}

\orcid{xxxx-xxxx-xxxx}
\author{Alicia (Hyun Jin) Lee}
\email{hlee3@andrew.cmu.edu}
\affiliation{%
    \institution{Carnegie Mellon University}
    \streetaddress{5000 Forbes Ave}
    \city{Pittsburgh}
    \state{PA}
    \country{USA}
}

\orcid{xxxx-xxxx-xxxx}
\author{Mai Lee Chang}
\email{maileec@andrew.cmu.edu}
\affiliation{%
    \institution{Carnegie Mellon University}
    \streetaddress{5000 Forbes Ave}
    \city{Pittsburgh}
    \state{PA}
    \country{USA}
}

\orcid{xxxx-xxxx-xxxx}
\author{Sreehana Mandava}
\email{sreehanamandava@utexas.edu}
\affiliation{%
    \institution{University of Texas at Austin}
    \streetaddress{}
    \city{Austin}
    \state{Texas}
    \country{USA}
}

\orcid{xxxx-xxxx-xxxx}
\author{Destiny Deshields}
\email{d240@umbc.edu}
\affiliation{%
    \institution{University of Maryland}
    \streetaddress{}
    \city{Baltimore County}
    \state{Maryland}
    \country{USA}
}

\orcid{xxxx-xxxx-xxxx}
\author{Hugo Simão}
\email{hsimao@andrew.cmu.edu}
\affiliation{%
    \institution{Carnegie Mellon University}
    \streetaddress{5000 Forbes Ave}
    \city{Pittsburgh}
    \state{PA}
    \country{USA}
}

\orcid{xxxx-xxxx-xxxx}
\author{Aaron Steinfeld}
\email{steinfeld@andrew.cmu.edu}
\affiliation{%
    \institution{Carnegie Mellon University}
    \streetaddress{5000 Forbes Ave}
    \city{Pittsburgh}
    \state{PA}
    \country{USA}
}

\orcid{xxxx-xxxx-xxxx}
\author{Jodi Forlizzi}
\email{forlizzi@andrew.cmu.edu}
\affiliation{%
    \institution{Carnegie Mellon University}
    \streetaddress{5000 Forbes Ave}
    \city{Pittsburgh}
    \state{PA}
    \country{USA}
}

\orcid{xxxx-xxxx-xxxx}
\author{John Zimmerman}
\email{johnz@andrew.cmu.edu}
\affiliation{%
    \institution{Carnegie Mellon University}
    \streetaddress{5000 Forbes Ave}
    \city{Pittsburgh}
    \state{PA}
    \country{USA}
}

\renewcommand{\shortauthors}{Alicia Lee et al.}

\begin{abstract}
The increasing number of older adults who experience cognitive decline places a burden on informal caregivers, whose support with tasks of daily living determines whether older adults can remain in their homes. To explore how agents might help lower-SES older adults to age-in-place, we interviewed ten pairs of older adults experiencing cognitive decline and their informal caregivers. We explored how they coordinate care, manage burdens, and sustain autonomy and privacy. Older adults exercised control by delegating tasks to specific caregivers, keeping information about all the care they received from their adult children. Many abandoned some tasks of daily living, lowering their quality of life to ease caregiver burden. One effective strategy, \textit{piggybacking}, uses spontaneous overlaps in errands to get more work done with less caregiver effort. This raises the questions: (i) Can agents help with piggyback coordination? (ii) Would it keep older adults in their homes longer, while not increasing caregiver burden?
\end{abstract}




\maketitle

\section{Introduction}
The number of older adults in the US is predicted to double by 2050 ~\cite{alzheimer20152015, cdc2024alzheimers,ortman2014aging}. Older adults are also living longer, potentially doubling the number of Americans with dementia by 2060 ~\cite{alzheimer20152015, cdc2024alzheimers}. This rapid demographic shift creates a care gap. The U.S. needs more than a million additional direct care workers, a demand that far exceeds the current and predicted labor supply ~\cite{2023alzheimer_special}. This raises an urgent question: Who will provide care for all these older adults, and how will this burden be managed across society? The impact of this care gap is especially severe for people of lower-socioeconomic status (SES). They face a significantly higher risk of developing dementia and will likely have more challenges accessing human service providers or gaining entry to high-touch memory-care facilities ~\cite{statista_dementia_income_2023,wang2023socioeconomic,bodryzlova2023social,li2023associations,fabius2022associations, ou2024socioeconomic}. 

Informal caregivers (e.g., spouses, adult children, neighbors, friends) provide support that allows older adults to complete the Instrumental Activities of Daily Living (IADLs)~\cite{mynatt2001developing}. IADLs are the tasks most related to successfully remaining in one’s home (e.g., cooking, cleaning, and transportation). Informal caregivers who help older adults complete IADLs offer a win-win. First, older adults who wish to age in place win by remaining in their homes. Second, society wins by avoiding the expense of increasingly over-taxed care facilities. Some research even suggests that the informal caregivers win in that caring for loved ones offers emotional fulfillment~\cite{ratnayake2022aging, pew2015caregivers}. Unfortunately, asking informal caregivers to take on all of these care obligations is an imperfect solution. Informal caregivers frequently feel overwhelmed and experience burnout, especially when caring for both aging parents and their own children ~\cite{honda2025balancing, tolkacheva2011impact,  charenkova2023parenting,ahn2009impairment}. It seems unlikely they can scale their efforts to meet this growing challenge. 

Recently, there has been interest in agents as a potential solution to absorb some of the burden. Researchers suggest that agents could provide support for older adults, helping them complete IADLs and reducing informal caregivers' burden ~\cite{chang2025unremarkable, chang2024dynamic, zubatiy2021empowering}. This solution envisions agents as autonomous systems that sense the environment, have an understanding of the work that needs to be done, perform tasks, communicate with people or other systems, and are perceived to have basic social skills~\cite{chang2024dynamic, chang2025unremarkable}.

Prior research explored how technology can assist older adults and their informal caregivers. Researchers explored systems that monitor changes in health and that facilitate care coordination among informal caregivers. Other research had agents provide reminders, support daily routines, and offer social or cognitive assistance~\cite {hamilton2021augmented, pino2020humanoid,manca2021impact,stogl2019robot,kubota2022cognitively,astorga2022social,mathur2022collaborative}. Agents can assist informal caregivers by providing tools for task management, communication, and shared scheduling ~\cite{consolvo2004technology, hwang2020exploring, mynatt2001developing,zubatiy2021empowering}. These systems have shown success in improving visibility and task completion, reducing errors, and easing coordination. 

Such systems may be especially valuable for lower-SES older adults, who often face the greatest challenges. They frequently lack stable resources, digital access, or consistent caregiving support~\cite{veinot2018good}, and therefore may have the most to gain from supportive technologies.

At the same time, the benefits come with new responsibilities, such as troubleshooting, system setup, and data entry ~\cite{zainal2025exploring, vines2013making, kim2023roles}. In practice, it remains unclear whether such systems ultimately reduce the effort that currently overwhelms informal caregivers.

To investigate the feasibility of agent-supported care coordination for lower-SES older adults experiencing cognitive decline, we explored three research questions:
\begin{itemize}
    \item RQ1. How do lower-SES older adults experiencing cognitive decline and their informal caregivers coordinate care?
    \item RQ2. What strategies do they use to manage the burden, maintain independence, and navigate care relationships?
    \item RQ3. How might these practices inform the design of agent-supported care coordination systems?
\end{itemize}

To explore these questions, we conducted a repeated-interview study with lower-SES older adults who self-report experiencing early cognitive decline, along with one of their informal caregivers. We took a design perspective, centering the lived, relational, and situated experience of older adults and their informal caregivers. The work produced three main insights. First, we observed that older adults made slow, subtle compromises to the completion of their IADLs to avoid burdening their caregivers. Paradoxically, this strategy of surrender and delay meant to keep them in their home only seemed to accelerate the conditions that would force them out. Second, we observed older adults keeping secrets from their family caregivers to hide the amount of care they received. Finally, we identified piggybacking, the grouping of tasks to reduce the burden of completing IADLs, as a potential strategy agents might employ to generate more care for older adults with less effort from informal caregivers. 

This study represents the early phase of a larger design research project. This paper makes two contributions. First, it offers an empirical account of how lower-SES older adults experiencing early cognitive decline coordinate care with their informal caregivers. Second, it speculates on how agents that support piggybacking might effectively help older adults to remain in their homes longer. 

\section{Related Work}
Our research draws from prior Aging in Place research, the need to complete instrumental activities of daily living (IADLs) to remain at home, and the role informal caregivers play in helping older adults complete their IADLs. We then examine HCI research to support Aging in Place, including work on care coordination.

\subsection{Aging in Place, IADLs, and the Role of Informal Caregivers}
Aging in Place (AIP) refers to the personal and societal desire for older adults to remain in their homes longer before transitioning to institutional care ~\cite{nia2023aginginplace, ratnayake2022aging}. To successfully age in place, older adults must manage their IADLs. These include tasks such as preparing meals, managing finances and health, caring for the home, maintaining personal hygiene, arranging transportation, and sustaining social connections ~\cite{edemekong2019activities, guo2025instrumental}.

As cognitive or physical decline progresses, informal caregivers often assist with IADLs. Informal caregivers most often include spouses, family members (including adult children), neighbors, and nearby close friends~\cite{sharma2016gender}. They provide unpaid support. Informal caregivers frequently report emotional strain, physical fatigue, and financial pressure that comes from providing care ~\cite{tang2018awareness, consolvo2004technology}. In many families, a primary informal caregiver, often an adult child, emerges to coordinate and carry out the bulk of day-to-day caregiving responsibilities. They do most of the work and organize support from other members of the care network ~\cite{schurgin2021isolation, renyi2022uncovering, consolvo2004technology, tang2018awareness}. The primary informal caregiver role is particularly challenging for those in the “sandwich generation”, individuals who balance elder care with responsibilities for their own children, households, and careers ~\cite{honda2025balancing, tolkacheva2011impact, charenkova2023parenting, ahn2009impairment}. The compounding demands of caregiving and personal obligations can lead to significant investment of time as well as caregiver burnout ~\cite{ahn2009impairment}. When the caregiver burden becomes unsustainable, it forces older adults to transition from home to institutional care ~\cite{sun2021adaptation, gaugler2003caregiving, taylor2023supportive, lethin2016family, tate2023factors}.

\subsection{HCI systems to support Aging in Place and address IADLs}
HCI research explored how technology might support older adults in managing IADLs ~\cite{forlizzi2004assistive, gasteiger2021friends, hawkley2019us, kubota2022cognitively, mynatt2001developing, barg2017understanding}. Prior work introduced a range of systems, including medication reminders ~\cite{tiwari2011feasibility}, physical activities and fitness systems~\cite{caldeira2023compare, caldeira2017senior}, nutrition coaching tools ~\cite{el2020virtual, siewiorek2012architecture, mccoll2013meal}, smart furniture ~\cite{forlizzi2005sensechair}, and conversational health assistants ~\cite{brewer2022if}. These tools aim to assist with specific tasks, maintain the older adult’s functional independence, and reduce caregiver burden.

HCI work also focused on the coordination of care across networks of informal caregivers ~\cite{forlizzi2007robotic}. Systems such as shared digital calendars, messaging platforms, and dashboards have been developed to help informal caregivers communicate, track tasks, and plan appointments ~\cite{mynatt2001digital, tang2018awareness, li2023associations, caldeira2017senior, consolvo2004technology, vines2013making}. 

The research has not produced a magic bullet for completing more IADLs while decreasing caregiving burden. Some studies even suggest that support systems create more burden by introducing new responsibilities and routines for caregivers. Managing digital systems and resolving technical issues creates more informal caregiver burden ~\cite{vines2013making}. In reality, the responsibility of organizing and maintaining these systems tends to fall on the primary informal caregiver ~\cite{schurgin2021isolation, kim2023roles}.

\subsection{Cognitive Decline and Care Coordination}
There has been increasing focus on how to support the growing number of individuals with cognitive decline beyond typical aging~\cite{alzheimer20152015, cdc2024alzheimers}. Cognitive decline introduces unique disruptions to how care is requested, remembered, delivered, and coordinated. 

Older adults with cognitive decline face increasing difficulty managing IADLs. These include tasks like remembering whether they have taken medication, managing finances without making costly mistakes, maintaining their homes, getting transportation to appointments or the grocery store, and staying socially connected~\cite{jekel2015mild, feger2020incident}. When these IADLs are no longer reliably performed, the burden of caregiving rises quickly~\cite{jekel2015mild}. HCI research has explored a range of systems to help AIP with cognitive decline ~\cite{mathur2022collaborative, zubatiy2021empowering}. Systems have helped with personalized reminders of care plans ~\cite{hamilton2021augmented}, motor and memory training tools ~\cite{pino2020humanoid, manca2021impact, stogl2019robot}, conversational agents for social support, and agents that help with daily routines and medication management ~\cite{astorga2022social, brewer2022if}. 

These systems support informal caregivers by providing real-time status updates, distributing tasks, facilitating communication within a social network, and clarifying responsibilities. These systems are often built to improve transparency, reduce missed care, and support equitable responsibility sharing \cite{schurgin2021isolation, consolvo2004carenet, hwang2020exploring}. While such tools improve visibility and reduce errors, they typically assume there is a primary informal caregiver with the time and capacity to operate them. In reality, care networks for older adults often include multiple caregivers with varying degrees of involvement ~\cite{forlizzi2007robotic}. Designing technologies for such contexts requires more than coordination logic. It requires an understanding of how to navigate and support complex, interconnected relationships.

Recent work has begun to examine how agents might assist with AIP among older adults experiencing cognitive decline in socially sensitive care networks ~\cite{reig2021social}. Chang et al. proposed the concept of \textit{agent affiliation}—i.e., who the agent appears to work for. Their work highlights that trust in an agent is shaped by who it appears to serve: the older adult, their family, or an outside institution~\cite{chang2024dynamic}. Their findings suggest that affiliation may shift over time: early in cognitive decline, older adults expected agents to support their autonomy, but as decline progresses, they envisioned the agent gradually aligning with an informal caregiver to reduce burden and assist with difficult decisions. Related work also highlights the social tensions surrounding agent support. In another study, Chang et al. showed that agents may shift roles depending on the older adult’s level of decline. Agents begin in a passive, background role (low presence, low obligation) and gradually become more active—protective, persuasive, and advocacy-oriented—as the older adult becomes less able to self-advocate~\cite{chang2025unremarkable}.

\subsubsection{Lower-socioeconomic status Older Adults and Compounding Challenges}
Lower-SES older adults face significantly higher risk for dementia and cognitive decline ~\cite{bodryzlova2023social, wang2023socioeconomic, veinot2018good, statista_dementia_income_2023}. They tend to experience cognitive decline earlier, and it progresses more rapidly ~\cite{pew2015caregivers, cdc2024alzheimers, wang2023socioeconomic}. 

Environmental factors further compound these risks. For example, residents of lower-income neighborhoods often exhibit lower baseline cognitive function and steeper memory loss~\cite{pew2015caregivers}. Financial stress is also related to an increased dementia risk ~\cite{li2023associations, ou2024socioeconomic}.

Unfortunately, lower-SES older adults frequently face substantial barriers to accessing professional care~\cite{jones2024healthcare, yang2024digital}. Paid caregiving, medical services, and digital health tools are often unaffordable or unavailable~\cite{fabius2022associations}. 

Technology may hold promise for improving care coordination in resource-constrained settings. While many prior systems have effectively supported a dedicated informal caregiver~\cite{tang2018awareness, vines2013making, consolvo2004carenet, hwang2020exploring}, the benefits may not reach lower-SES families. In many cases, primary informal caregivers often relocate or leave full-time employment to provide sustained care ~\cite{consolvo2004carenet,tang2018awareness}. Such coping strategies are often infeasible for lower-SES families. Yet these families are often those who need such technologies the most. As a result, such systems may not always translate well to families with fewer resources, potentially leaving behind those with the greatest need~\cite{veinot2018good}.

\section{Method}
We employed a Research through Design approach to critically investigate the vision of agents supporting older adults experiencing cognitive decline to assist with AIP ~\cite{frayling1994research, zimmerman2007research}. We wanted to gain a sufficient understanding of the present to envision a preferred future that could and should be.

We chose to conduct repeated interviews for approximately three weeks with ten lower-SES older adults experiencing cognitive decline and one of their informal caregivers. Repeated interviews where we probed participants on how they coordinated their care, how their plans broke down, and how they strategized completion of their IADLs. Repeated interviews can help surface everyday adaptations that might not be captured in a single interview. By spacing the interviews across multiple weeks, we observed some changes over a short period of time and followed up on tasks in progress. The repeated encounters also allowed us to build trust with the participants.

\begin{table*}[ht]
\renewcommand{\arraystretch}{1.6} 
\setlength{\tabcolsep}{10pt}       

\begin{tabular}{ll|lll|lll}
\noalign{\hrule height 1.2pt}
\textbf{Pair} & \textbf{Relationship} & \textbf{ID} & \textbf{Age} & \textbf{Gender} & \textbf{ID} & \textbf{Age} & \textbf{Gender} \\
\noalign{\hrule height 1.2pt}

P1  & Friend / Neighbor        & OA1  & 81 & Female & ICG1  & 75 & Female \\ \hline
P2  & Spouse / Partner         & OA2  & 64 & Male   & ICG2  & 68 & Female \\ \hline
P3  & Grandparent / Grandchild & OA3  & 80 & Female & ICG3  & 19 & Male   \\ \hline
P4  & Friend / Neighbor        & OA4  & 83 & Female & ICG4  & 78 & Female \\ \hline
P5  & Spouse / Partner         & OA5  & 71 & Male   & ICG5* & 71 & Female \\ \hline
P6  & Friend / Neighbor        & OA6  & 76 & Female & ICG6* & 73 & Female \\ \hline
P7  & Parent / Adult Child     & OA7  & 83 & Female & ICG7* & 62 & Female \\ \hline
P8  & Friend / Neighbor        & OA8  & 67 & Female & ICG8* & 83 & Female \\ \hline
P9  & Spouse / Partner         & OA9  & 78 & Female & ICG9* & 77 & Male   \\ \hline
P10 & Friend / Neighbor        & OA10 & 86 & Female & ICG10 & 69 & Female \\

\noalign{\hrule height 1.2pt}
\end{tabular}
\vspace{1\baselineskip}
\caption{Participants’ demographics. Ten older adult–informal caregiver pairs participated (20 individuals). An asterisk (*) indicates informal caregivers who self-reported cognitive decline.}
\label{tab:table_demographic}
\end{table*}
\label{tab:table_demographic}

\subsection{Participants} 
We recruited participants through local senior centers and from high-rise residences that cater to older adults in low SES neighborhoods. We focused on participants from low SES backgrounds, as they experience higher rates of dementia~\cite{2023alzheimer_special}. We qualify participants as lower-SES if they met one or more of the following criteria: zip code (and associated housing costs), eligibility for local housing support programs, pre-retirement occupation, level of education, and annual income. Typically, organizations categorize lower occupational status by various factors, such as educational requirements, average income levels, and the skill or expertise required ~\cite{hollingshead1970commentary, organisation2010pisa}. 

We collected data from 10 older adult-informal caregiver pairs (20 participants). Older adults whose ages ranged from 64 to 86, and informal caregivers whose ages ranged from 19 to 83. Sixteen participants identified as female and four as male. Eligible older adults were those who self-reported experiencing cognitive decline beyond typical aging and who had at least one informal caregiver. One participant has a diagnosis of MCI. Five informal caregivers reported experiencing cognitive decline beyond typical aging at times. All participants were capable of providing informed consent. They all understood the goals and activities of our research study. Table \ref{tab:table_demographic} summarizes participant demographics.  

\subsection{Procedure}
Our study was approved by our University’s Institutional Review Board. As part of our ethical protocol, participants were informed about the study’s goals and the concept of an agent meant to help them AIP. We covered this during recruitment, prior to signing the consent form, before the first interview, and again before the second and third interviews. 

Each pair of older adults and informal caregivers participated in three semi-structured interviews, spaced approximately one week apart. This allowed us to follow up on emerging themes, observe changes in plans or circumstances, and deepen rapport. The first session explored routines and support networks, and the second one revisited and reflected on specific care interactions and challenges, including how care was negotiated. The final session adapted to emerging themes to probe deeper into each pair’s situation. This structure allowed the research team to thoroughly assess their physical environment and the effectiveness of their current strategies for completing IADLs. Researchers worked in teams of two to three. One researcher conducted the interview, while the others observed and took notes.

For the first session, which lasted about 90 minutes, researchers visited the pair at the home of either the older adult or their chosen caregiver and got to know them. After obtaining informed consent and demographic information, the semi-structured interview explored participants’ daily and weekly routines and their support networks. We asked about the relationships between older adults and informal caregivers, their respective roles and responsibilities, and how their caregiving relationship began. Participants walked us through how they planned and carried out routines across different days of the week, focusing on IADL-related tasks and who provided support for them. Interviewers documented these routines through field notes and photographs of coordination tools such as calendars, phone memos, to-do lists, notebooks, and planners. At the end of the interview, participants were asked to share their plans for the upcoming week. 

One week later, we conducted a second interview, which lasted about 30 minutes. We revisited the care plans discussed in the previous session and asked whether and how those plans were carried out. If plans were not completed, we discussed the cause of the breakdowns. These conversations helped us identify disruptions, understand participants’ coping strategies, and observe how coordination adapted over time. We also explored ongoing challenges, desires, and needs related to care.

The third interview, also 30 minutes, was conducted one week after the second interview. It was built on insights from the previous two sessions. We adjusted our questions to probe deeper into each pair’s situation. For example, if a friend was identified as the primary driver, we asked what would happen if that friend were unavailable. If transportation was a recurring issue, we explored how participants handled last-minute ride cancellations or changes in availability. After each round of sessions, the research team met to refine questions for the following week, allowing the interviews to be tailored to emerging themes.

To better gain empathy for our participants and the lived experience of being a lower-SES older adult, one researcher joined a local senior crochet group. They attended, meeting weekly for six months, spending 3–4 hours per session with a small group of 6–8 older adults. They also volunteered at the senior center, where we did our recruiting. This provided opportunities for informal, in-depth conversations and observations. The research team also had conversations with diverse professionals, including a psychologist, a geriatrician, and staff at senior centers. Through these activities, we gained insight into some of the challenges older adults face.

\begin{figure*}[ht]
    \centering
    \includegraphics[width=0.45\linewidth]{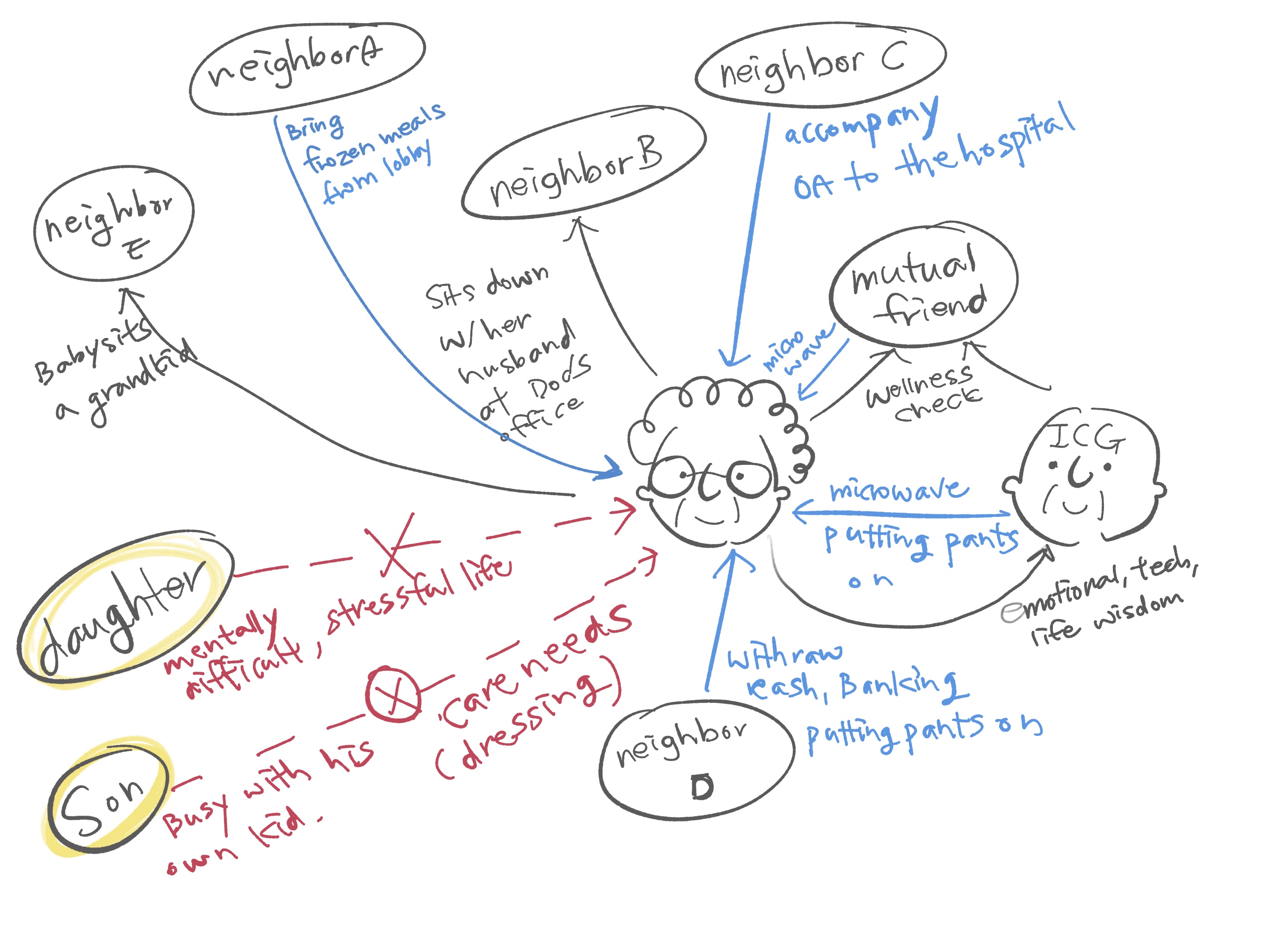}
     \includegraphics[width=0.45\linewidth]{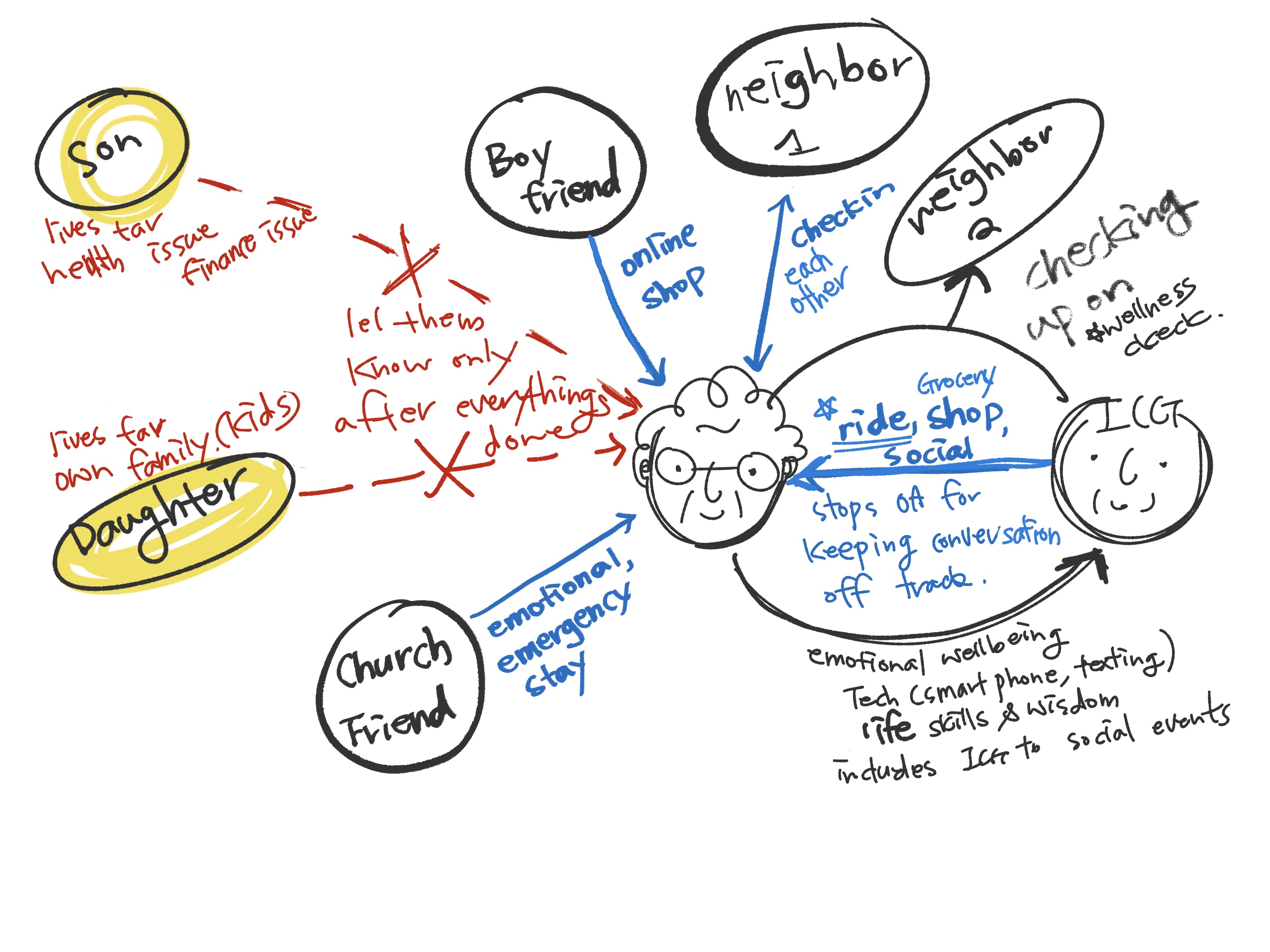}
    \caption{Blue arrows indicate the support the older adult receives (e.g., clinic sit-ins, rides, meals, dressing help, cash/banking). Black arrows indicate the support the older adult provides (e.g., babysitting, accompanying others, emotional check-ins). Yellow family nodes show limited or withheld involvement, while nearby peers coordinate frequently through small, lightweight tasks.}
    \Description{Hand-drawn relationship map with an older adult at the center. Blue arrows point inward from neighbors and an ICG to tasks they provide (hospital rides, dressing, banking). Black arrows point outward from the older adult to tasks they provide (babysitting, accompanying). Family nodes are present but minimally connected; peer nodes are densely connected.}
    \label{fig:relationship_map}
\end{figure*}

\subsection{Analysis}
All interviews were audio-recorded and automatically transcribed. Researchers wrote field notes capturing their observations from each interview. We replayed the recorded interviews, reviewed the field notes, and reviewed the transcripts. Using our research questions as a guide, we documented observations and insights on sticky notes and created affinity diagrams. We built and rebuilt clusters of sticky notes to identify recurring themes, coordination and coping strategies, and challenges across participant pairs~\cite{ixdf_conceptual_models}. In addition to thematic clustering, we sketched conceptual models that capture an abstraction of the current state (\autoref{fig:relationship_map}). This helped synthesize information about the people involved, the types of tasks exchanged, and the directionality of support. These models helped us visualize distributed care structures, role fluidity, and ad-hoc reciprocity within each network. Similar approaches have been used in prior HCI research to synthesize data and reveal patterns in complex, socially situated practices~\cite{chang2025unremarkable, chang2024dynamic, luria2020social, luria2019re, reig2020not, karapanos2009user, zimmerman2008role}. We held discussion sessions among the research team to interpret and refine our findings. Team discussions were particularly valuable for uncovering social dynamics that were not always explicit in participants’ utterances or in our field notes. 

We chose to use affinity diagrams to analyze our data (instead of thematic analysis or grounded theory) because of our design focus and RtD approach. We were not trying to develop a theory of human behavior. Instead, we wanted to discover actionable insights that might improve the design of agents. This paper represents early work in a larger RtD project that will be informed by this study. This sequence of first publishing on field work and then on artifact creation and assessment is well established in HCI design research~\cite{wakkary2007resourcefulness, odom2011teenagers, yoo2013bus}. For example, Odom et al.~\cite{odom2011teenagers} first conducted fieldwork exploring how teens engage with their material and visual possessions before building a super teenage bedroom of the future where they could blend their virtual and material things. Similarly, Yoo et al. ~\cite{yoo2013bus} conducted a field study on transit riders before building a crowd-sourced real-time arrival system ~\textit{Tiramisu}.

\section{Findings}
Our analysis surfaced three interconnected insights related to care coordination and the completion of IADLs: 
\begin{enumerate}
    \item Older adults continually accepted lowering standards of living to reduce complexity and to ease informal caregivers' burdens.
    \item Care coordination involved small, reciprocal exchanges, spread across a broad network of informal caregivers and shaped by secrets kept from specific informal caregivers about certain kinds of support. 
    \item Older adults and informal caregivers liked to use piggybacking, a strategy that matched an older adult care task with a task an informal caregiver completed for themselves. This reduced the burden for completing the work.
\end{enumerate}

These practices allowed participants to maintain a self-presentation of independence; minimized the visibility of their actual need for care; and avoided overburdening informal caregivers. 

Our notation is as follows: for older adults, we use participant number, gender (“M” for male, “F” for female), and age (e.g., OA1-F87). For informal caregivers, we use participant number, gender, and age (e.g., ICG1-M27). See \autoref{tab:table_demographic} for details.

\subsection{Decaying Kingdoms}
In contrast to previous work~\cite{kim2023roles,hwang2020exploring,schurgin2021isolation}, older adults in our study functioned as the primary coordinators of their own care. Many appeared to be losing their ability to manage IADLs. They frequently made small compromises that reduced the overall work getting done. They slowly ceased activities like maintaining their home, attending appointments (including doctors’ appointments), and preparing healthy food. This seemed to be motivated by a desire to prevent a growing burden on informal caregivers and to preserve an appearance of independence. Interestingly, this strategy only seemed to accelerate the chances of having to move, as IADLs were not being addressed. 
\begin{figure*}[htbp]
    \centering
    \vspace{4\baselineskip}
    \includegraphics[width=1\linewidth]{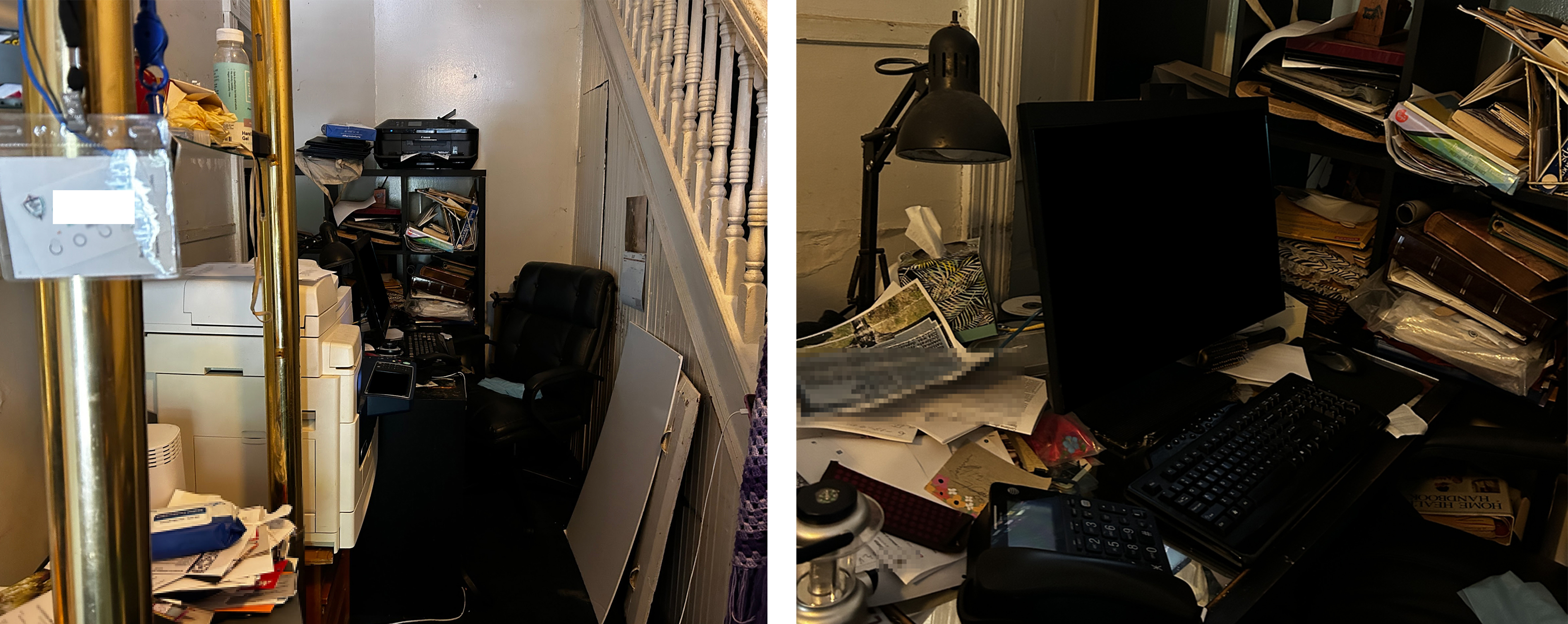}
    \caption{‘Command station’ under the narrow staircase.}
    \Description{Photo of a small desk placed under a narrow staircase. The desk piled with piles of medical paperwork, bills, other papers, pens, and a computer. A storage unit next to the desk blocked half of the passage between the entryway and the living room. The storage unit held loose papers and books stacked haphazardly.}
    \vspace{2\baselineskip}
    \centering
    \label{fig:desk}
\end{figure*}
\begin{figure*}[htbp]
    \centering    
    \includegraphics[width=1\linewidth]{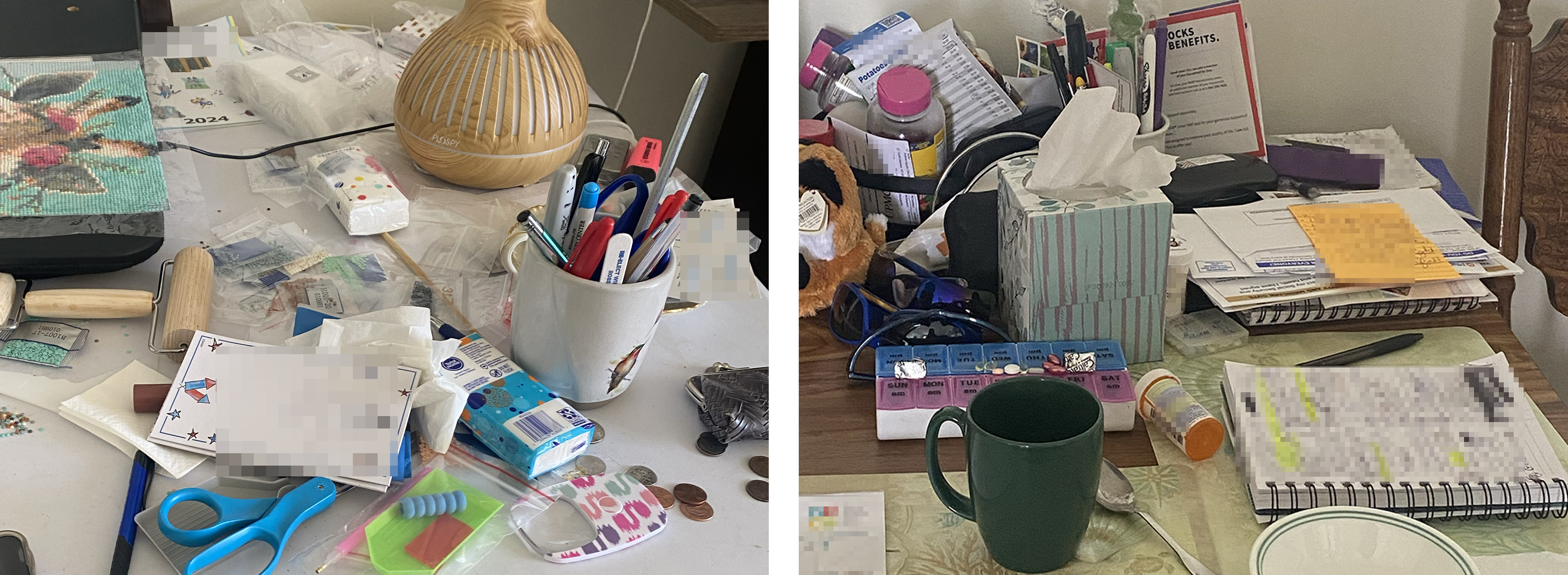}
    \caption{Cluttered dining surfaces. Items were intermixed on tables that were also used for meals. Older adults could easily direct informal caregivers to them when needed, which appeared to reduce the effort for informal caregivers to search through the home or enter private areas like the bedroom.}
    \Description{Photo of a dining table used for both crafts and meals; materials and dishes are visible on the same surface, indicating a shared, multi-purpose workspace.}
    \label{fig:table}
    \vspace{4\baselineskip}
\end{figure*}

Evidence of this strategy revealed itself in the state of their homes. The rooms were dangerously full, layered with objects accumulated over decades. To minimize the effort involved in accessing items needed for daily tasks, their critical items (e.g., multiple remotes, checkbooks, cash, medication, important documents) formed a precarious top layer. Similar to prior work, all of the homes we visited held loosely arranged, critical items within arm’s reach of where the older adult spent most of their time ~\cite{forlizzi2005sensechair, forlizzi2004assistive}. These zones were typically centered around recliners or couches. Due to the compact size of their homes and apartments, these zones were often located close to the entryways. We reasoned that these physical spaces were signals of losing control and, at the same time, minimizing the need to ask for help. Prior work noted that cluttered environments complicate health management ~\cite{hirsch2000elder}. Our participants seemed to tolerate and even normalize the clutter. This acceptance reflected a broader pattern of incremental compromise. They seemed to accept clutter as a way to avoid drawing on caregiver support and to preserve a sense of independence (See~\autoref{fig:table}). 

Many participants shared why they had abandoned previous storage strategies. Reasons included the weight of items, cramped spaces, or simply the convenience of placing items close to where they were used (e.g., floor or doorway). Participants kept items visible to function as a reminder of what they needed to do and to avoid misplacing or forgetting where something was recently stored. While they once placed items on shelves or in cabinets, these strategies for maintaining a usable home were abandoned as the effort of reaching high places or bending to access low storage became more difficult. Interestingly, they said they did not ask their informal caregivers to help them with putting things away or to help them develop new storage strategies that placed high-use items within easier reach. Instead, they stacked items on countertops, table tops, or other flat surfaces. As these spaces became overrun, they began stacking items directly on the floor. Reorganizing items felt like an overwhelming amount of work, given the current state. Piles of items formed narrow paths through rooms. They increased the chance of tripping and made traversing their homes increasingly more difficult.

As an example, OA4-F83’s living room was densely packed. A layer had formed at waist-height. The majority of items were rarely used except for the thin top layer. We had to step around a couch that partially blocked the entryway. Moving around the room was difficult and dangerous. If we brushed against the piles, we risked starting an avalanche. When our participant fetched a drink, we saw them push the couch aside and shuffle through the narrow gap.

Behind the couch, a glass cabinet displayed sentimental artifacts. The doors could not be opened because the couch blocked their movement. Inside were long-dead Mother’s Day flowers, beaded crafts, and feather ornaments. Under a narrow staircase, the participant had established a command station: a desk piled with piles of medical paperwork, bills, other papers, pens, and a computer (see~\autoref{fig:desk}). 
A storage unit next to the desk blocked half of the passage between the entryway and the living room. The storage unit held loose papers and books stacked haphazardly. To the right of the entryway, bulk goods and canned food were stacked in their packaging directly on the floor. This participant no longer seemed to place new items coming into their home in closets or kitchen cabinets. 

Many participants stored groceries and bulk items near the entryway, within easy reach. In homes with staircases, the stairs often served as additional storage. Items frequently lining each step, tucked against the wall. When we inquired about how they used the stairs, almost all participants shared that the stairs were only used twice each day: once to come downstairs in the morning and once to go upstairs at night to sleep. 

Descriptions of their daily routines revealed additional clues about loss of control around IADLs. Many no longer cooked. Participants often skipped meals, eating irregularly and only when they felt like eating. No one mentioned being hungry. They relied on prepared canned goods (e.g., soup), frozen dinners, or meal delivery from assistance programs. Others would dine out and then save leftovers, stretching these across multiple days. Breakfast cereals and snacks like cookies and chips as meal replacements. A few made simple dishes such as boiled potatoes. As OA10-F86 shared, \textit{“I'm a flexible person. I do not eat meals at certain times... I might have something left over, or I might eat another Meals on Wheels.”} Participants explained that cooking for one felt like too much effort. They complained that fresh ingredients spoiled too fast. They no longer made weekly trips to the grocery store. 

The surrender of IADLs was mirrored in other aspects of participants’ lives. Many postponed bathing because of the effort and a fear of falling. Appointments were often canceled when the effort to arrange transportation or their energy level for the day was low. They found cancellation to be an easy option. They noted that they did not worry much about breakdowns, like a ride falling through, because activities could simply be canceled.

Participants presented themselves as satisfied, competent, and in charge. However, their physical environments, poor diets, declining health management, and habit of canceling activities and appointments told a more complicated story. These descriptions suggested a continual acceptance of small defeats and compromises that negatively impacted completion of IADLs. Participants seemed to be surviving much more than thriving. Any adverse event, such as a fall, might force them to leave their home. They did not ask for help, and they seemed to be in a bit of denial about how precarious their lives had become. None of the informal caregivers commented on the many things that were not getting done. Informal caregivers did not speak of the older adult as resisting their help to improve living conditions.

\subsubsection{Relational Dynamics of Informal Care}
Family ties prioritized role-preservation, and peer ties deepened into interdependence. When multiple helpers were available, older adults matched requests to the task and the relationship.

Family caregivers and older adults navigated pride, frustration, protectiveness, worry, and affection. They worked hard to help older adults remain the spouse, parent they had always been. This dynamic often required caregivers to absorb the burden of errors quietly. OA02-M64 was proud of his home management skills. He walked us through Siri alarms, digital calendar entries, and scrolled through dozens of notes. Across from him, his wife (ICG02-F68) turned to the researcher and mouthed \textit{“no.”} Later, she leaned in and whispered that essentials were often missed. She let him keep the role anyway and quietly picked up what he had forgotten. The compromise cost time and energy but preserved who he had been. 

Peer caregiving often grew from neighborly ties into interdependence. OA4-F83 and ICG4-F78 had been next-door neighbors who nodded at the grocery store for years. They began seeing each other more while caring for a mutual friend, and similar life events drew them closer. Every morning, one called the other to share plans. 

Many participants emphasized remaining in their community and avoiding burdening adult children. Frequent, low-effort contact sustained support. Most days, neighbors met by the elevator, compared notes on who seemed unwell on the floor, and assembled short rotations for wellness checks when needed. With more than one caregiver, older adults appeared to sort requests and route them accordingly. Quick tasks (e.g., microwaving a frozen meal) went to hallway neighbors who were available in the moment; transport needs went to a friend with a car already headed nearby; tasks that might compromise a sense of independence or image (e.g., dressing) were reserved for next-door peers rather than adult children. These were not fixed rules. They shifted with day-to-day availability, perceived burden, and task type. What was shared also varied. For example, one older adult shared sensitive personal information with neighbors who helped with bank errands. The older adult left checks, a note of personal details, and cash on a table by the front door so helpers did not need to enter the bedroom.

\subsection{Can You Keep a Secret? Care Coordination is Selective}
Older adults controlled when and how they received support, in terms of who knew about the different kinds of support they received.  They asked for help across a large and loose network of friends and neighbors, making small requests of a large number of people in their networks. They also discovered opportunities to create reciprocity. They wanted to do something for the people who helped them. These findings extend previous work showing how such networks are activated through a series of small negotiations and opportunistic requests that preserve the older adult’s sense of control ~\cite{forlizzi2004assistive}. These strategies, when asking for help, prioritized the preservation of their autonomy and independence over completion of all IADLs. Their need to control their care was reflected in their fear of having to move. Many strongly asserted that as long as they could manage their daily life on their own, there was no need to leave their current home.

\subsubsection{Many Small Acts of Receiving and Giving Care}
Almost all of the older adults’ care requests could easily be completed during a single, short visit. When tasks ended up being larger than expected (e.g., more than an hour, more than one visit),  older adults generally delayed or canceled them. Older adults avoided requests that might seem large or feel burdensome, such as cooking an entire dinner or carrying many heavy grocery bags into their home. They worked to break work down into smaller tasks that could be divided across their network of informal caregivers. For example, OA10-F86 described a chain of support involved in her meal preparation, \textit{“My Meals on Wheels are delivered [frozen]. Because it's just getting harder for me with my hand– I have hand problems– so my neighbor will bring the meal [up from the lobby]. And anything goes in the microwave, if I need to heat up something– it's harder for me, so my other neighbor will come over and get it out because I don't have the strength in my hands to do that.”}

Many participants reinforced the importance of giving back by helping those in their community, as opposed to simply requesting support, which implies decline and dependency. OA1-F81 received periodic deliveries of fresh produce from a low-income food assistance program. She mentioned that her neighbor, who supported her with various tasks, often took some of the fresh items that the older adult did not need or want. This reinforced a sense of mutual benefit. 

Older adults also provided support to friends and informal caregivers when the opportunity arose, reinforcing their position as active contributors within the network. For instance, one older adult regularly received support with tasks such as grocery pickups or help getting dressed from nearby informal caregivers. In return, she watched one informal caregiver’s husband, who had a health condition that required monitoring, while the informal caregiver ran errands. The OA10-F86 explained: \textit{“I think that's our attitude here… One of our neighbors has a little grand-nephew. If she needs to go someplace, she will leave him with me while she just runs and does something quick. I mean, that's just sort of a community that we have. One of my friends, she's the one that goes to the bank for me, when she needed to go someplace, I went down and sat with her husband while she was gone. Another friend’s daughter cooks on the weekends, and her daughter will send home stuff, not just for her mom, but for me too.”}

\subsubsection{Selective Disclosure to Manage Self-Presentation}
Participants selectively disclosed their needs, keeping secrets to control others’ views of their overall well-being. They did not want their informal caregivers (especially their adult children) to know how much care they were receiving. They shared only the information necessary for a task. For example, when a neighbor provided a ride to an appointment, they were typically told only the time and destination. The older adults did not share the reason for the visit, who else was helping, or what other care the older adult received. This “need-to-know” approach also kept each interaction lightweight. As ICG8-F83 put it, \textit{“I think that there could be all kinds of things about people just wanting to look a certain way in front of other people. But I also think that there's a level of it... I don't want to see that I'm declining. I don't want to recognize that I may have more needs than I did a month ago.”}

This approach extended to family relationships. Many participants intentionally hid signs of vulnerability from their adult children while disclosing the same needs to neighbors or friends. OA10-F86 used a power wheelchair and required daily dressing support due to a catheter. The informal caregiver, who lived next door, came over in the morning to help her put on her pants. The older adult mentioned that a couple more friends also helped her with dressing. This both spread the work out across her network and provided robustness for days when her neighbor might not be available or feel up to the task. To receive this critical support, the older adult kept her front door unlocked. She traded her security for critical care. 
The older adult actively worked to prevent her son from knowing about this arrangement. She shared the reason why, \textit{“My son, he works full time. He's got three kids that are now college age... I try not to call him any more than I have to. He gets mad sometimes because I don't call him often enough. But I feel like as long as I can be independent, not have to depend on him. If I can do it myself, I've always been a very independent person, and I plan to try to do that forever. So sometimes he'll give me the devil, ‘Well, mom, I could have done that easier!’ I don't like to do that.”}

Older adults worked to present an image of independence, particularly in the eyes of their children. This approach challenges the dominant design assumption in prior systems that informal caregivers need transparency and that transparency is essential in care coordination~\cite{schurgin2021isolation, consolvo2004carenet, hwang2020exploring,kim2023roles}. Our findings suggest that requests for transparency might not work. Older adults valued control over visibility more than efficiency. This reframes coordination as a relational practice sustained through micro-interactions, mutual aid, and the strategic management of information.
\subsection{Piggybacking as a Win-Win Strategy}
Piggybacking emerged as a low-effort, highly effective care strategy. It involved adding a small favor or task to an existing errand or trip that an informal caregiver already planned. For example, if an informal caregiver was already going to a pharmacy, then getting a prescription for the older adult from the same pharmacy provided assistance without much effort. Similarly, one informal caregiver was already going to a grocery store, then getting items for multiple older adults from a single trip. Piggybacking functioned as a win-win: it reduced the informal caregivers' effort by bundling support into existing routines, while aligning with older adults’ desire not to be a burden.

Piggybacking often emerged through situational strategies. These were shaped by how older adults responded to their informal caregivers’ routines: (1) building on known routines, and (2) co-discovering opportunities during social interactions. These opportunities encompassed a wide breadth of tasks, including getting items, asking for a ride, or assistance with an appointment. Opportunities were also typically flexible. Informal caregivers described piggyback requests as “nothing.” They did not create a feeling of being burdened. 

Piggybacking requests could build on the informal caregiver's known routines. For example, OA9-F78 called ICG9-M77 during his bike ride, asking him to pick up a specific grocery item from a store on his regular route. Another OA4-F83, sometimes requested a friend to withdraw or deposit money on her grocery trip, since the store was near the bank. These examples demonstrate how older adults' awareness of informal caregivers' routines in their neighborhoods could be leveraged for piggybacking requests. 

Piggybacking also arose from informal social interactions, such as hearing about an upcoming shopping trip while chatting on the phone. ICG4-F78, noted, \textit{“I said, I'm going to the bank. I'm going to get my prescription. The older adult will end up saying, ‘Oh, I’ll go. I’ll tag along.’”}  
During a hallway conversation, an older adult might mention needing bananas, and the friend would offer to pick them up while doing their own shopping. 

OA1-F81, who only drove to nearby locations, shared a similar example of spontaneous piggybacking. After she ran into her neighbor in the lobby, she recalled: \textit{“My neighbor said she was going to get her vouchers at [social service center W]. I said, ‘I’m going to [senior center, which is nearby W]. Do you want to go with me?’ She said, ‘Oh, sure.’ So we went together, and afterwards I said, ‘I got to stop at Aldi’s [a grocery store] for a second.’ She said, ‘Oh, stop anywhere you want, and I want to get something too.’”}

Requests could also emerge while older adults and informal caregivers were already together, where they negotiated what could be done jointly. These moments created natural ad-hoc coordination points. After events at the senior center, older adults and informal caregivers would decide on the spot whether to run errands together or to head home instead. For example, OA4-F83 received rides from a friend experiencing cognitive decline. They spoke almost daily and learned each other’s schedules. The older adult often joined the friend’s pharmacy trips to use coupons for milk or frozen meals. 

These piggybacking practices demonstrate how older adults actively seek opportunities that seem to be less demanding of informal caregivers and that preserve their sense of independence. By embedding care in small, opportunistic exchanges, older adults maintained an image of independence to their children. This reframes care as a social negotiation grounded in preserving agency. However, it is unclear if older adults will continue to have success creating piggybacking opportunities as their cognitive capabilities decline. 

\section{Discussion}
Our goal is to support older adults to remain in their homes by ensuring that IADLs continue to be completed, even as older adults face increasing cognitive decline. However, most IADLs are physical tasks that are currently performed by people (e.g., bathing, cutting toenails, driving to appointments, cleaning bathrooms, preparing food, washing dishes). Without a physical body, agents cannot easily help with much of the work. They can help to coordinate and optimize the use of available human resources, particularly by organizing informal caregivers to accomplish these tasks. Getting all the IADLs done so they can remain in their home is particularly challenging for lower-SES older adults. They lack stable access to paid caregiving, medical support, and a robust digital infrastructure~\cite{fabius2022associations,yang2024digital,jones2024healthcare}. 

Care coordination becomes complicated by mismatches between the kind of assistance that can be provided, what older adults need, and what older adults want their caregivers to know. Many older adults want to resist or deny the reality of decline. As a result, systems that emphasize efficiency, visibility, or central control may feel misaligned with their values. This differs from systems that respond to selective requests from older adults, which often happen in the moment. Appropriate framing is needed to design systems that help to complete more IADLs for older adults and that they will find acceptable.

Our study produced two main findings. First, older adults are not in control of IADLs, but they exercise strong control over the care that they request and receive, through what requests are made, who the requests are made to, and how much is disclosed about this coordination. Second, we identified piggybacking as an observed strategy in which more tasks were completed with less burden. Piggybacking actions — which we identified as strategic, social, and emergent — offered a lightweight, opportunistic form of coordination. These actions could be leveraged by agents to improve care outcomes while respecting the nuanced social dynamics of care. 

\subsection{Control Over Disclosure}
Even though many of our participants appeared to be in denial about their decline, they exercised strategic control over how and when to involve others in their care. older adults in our study carefully decided what to request from each informal caregiver and curated who knew what regarding their care and condition. This meant that some caregivers were selectively informed, while others were kept in the dark about particular needs. We inferred that these actions represented a deliberate effort to protect autonomy, manage social relationships, and avoid burdening others. 

Agent support should therefore not assume that full transparency is always beneficial for older adults. Prior work has shown that informal caregivers often want more transparency around caregiving for older adults, and that transparency can indeed help with coordination ~\cite{schurgin2021isolation, consolvo2004carenet,hwang2020exploring}. However, our findings reveal that older adults often resist this level of visibility. For them, maintaining control over disclosure is a crucial way to preserve a sense of control even in the face of decline. 

This tension raises open questions. Can systems help identify what is being neglected without requiring full disclosure from older adults? Can agents work effectively when only fragments of information are shared with informal caregivers? Addressing these questions will be important for developing systems that both respect older adults’ values and improve the completion of everyday tasks. 

\subsection{Piggybacking: Doing More with Less}
Piggybacking emerged as a successful strategy in our study, minimizing both physical effort and social cost for older adults and informal caregivers. These opportunities were often strategically identified, arose during informal social interactions, and were emergent. Piggybacking thus made coordination feel lightweight, natural, and opportunistic rather than burdensome. These findings highlight the need to shift the design challenge. Alongside building smarter tools, we must also enable more strategic and context-aware coordination within existing human and social infrastructure.

While effective, piggybacking also has limitations. It does not always reflect the older adult’s actual priorities, since requests that follow the schedules and errands of informal caregivers may prioritize other things. Opportunities for piggybacking are often hidden until they surface anecdotally through chance encounters. As a result, many potential overlaps in routines remain undiscovered. 

Agents can help reduce these limitations by actively surfacing opportunities for additional piggybacking. An agent could detect overlapping needs, routines, and resources, surface lightweight piggybacking opportunities, and recommend tasks to available informal caregivers, reducing burden and improving IADL completion.

However, care and dignity must be prioritized with every caregiving act. This approach may be especially relevant for lower-SES older adults in the absence of formal supports~\cite{jones2024healthcare,fabius2022associations}. It is important to design agents that can identify and support socially acceptable forms of piggybacking without causing tension in social relationships.

This type of piggybacking agent raises questions for further consideration. It is important to consider how accurately agents can model and detect piggybacking opportunities that older adults strategically identify or that are identified through casual social interaction. It is also necessary to consider how agents can predict the potential for piggybacking, some of which is emergent in nature.

\subsection{Secrets, Denial, and Blind Spots}
Our observations suggest that both older adults and caregivers create “blind spots” that delay help. This makes early noticing a promising design opportunity, though it raises privacy questions.

Older adults appeared to downplay or ignore their reduction in physical and cognitive capabilities, presumably because they were either in denial or they wanted to give an impression of wellness and independence~\cite{chang2025unremarkable}. We also observed that informal caregivers did not discuss the state of the older adults they cared for. We inferred that they preferred not to confront older adults' increasing decline out of sensitivity. Some may also fail to notice gradual changes because they see older adults nearly every day. Collectively, these behaviors created blind spots where the opportunity to provide care or to help sustain an older adult's wellness and independence went unnoticed.

These blind spots could be addressed by agents acting as an  “early noticer” before a situation becomes unmanageable. A system could gently flag missed tasks, such as clutter that accumulates in a space based on multi-modal inputs (e.g., visual cues from video, motion inactivity, or changes in routines). For lower-SES older adults, such blind spots may remain unnoticed for longer periods, due to limited access to regular medical check-ins or paid caregiving oversight~\cite{jones2024healthcare, fabius2022associations,yang2024digital}. 

Future work must examine whether agents can intervene in subtle, non-threatening ways during pre-critical stages of decline. For example, agents can use relationship-aware information sharing. In early stages, send a minimal, actionable task detail to a nearby peer who can help now; later, send a brief summary to distant family to provide peace of mind. 
It is also important to understand how systems can distinguish when intervention is helpful versus when it becomes intrusive. These questions may require perspectives from clinicians and elder law experts.

\section{Conclusion}
This paper reports findings from interviews with older adults and their informal caregivers, examining how care is coordinated under conditions of cognitive decline and limited resources. Older adults experienced cognitive and physical decline and seemed unable to stay on top of IADLs. Instead, they maintained control over who knew what about their situation and from whom they requested assistance. Piggybacking, the grouping of small, finite tasks together by one caregiver, was a mechanism that helped older adults and informal caregivers accomplish more with less. Piggybacking requests were strategic, in line with the control mentioned above, often emerging through informal social interaction or other circumstances. Based on these findings and insights, we present design considerations and future research directions. These will help guide the development of agents that support AIP while respecting the values and social dynamics of older adults and informal caregivers.

The study forms the early phase of a larger design research project. Its insights establish the qualitative foundation for subsequent prototyping and evaluation with stakeholders in the care network.

\section{Limitations}
We acknowledge several limitations in our participant sample and approach. Although we had a higher number of female participants, this aligns with demographic patterns of females both live longer and are more likely to take on informal caregiving roles~\cite{cdc_life_expectancy_2023, us_census_older_population_2020,sharma2016gender}. 

Participants self-reported experiencing cognitive decline. None had a formal dementia diagnosis. Therefore, the findings may not represent the coordination challenges in more advanced cognitive impairment.

We relied on verbal interviews and researcher observations. Across repeated interactions, all participants were verbal and capable of providing informed consent, validating the reliability of their accounts. Even so, future work could enrich these data with co-design and other participatory methods.

This study aimed to understand what to make and why, in depth. We did not implement sensing or prompting mechanisms. Future work should test these capabilities to examine how agents can surface opportunities in practice. 

Our findings are exploratory and involved a small sample in the United States for a short period of time. They provide depth but not statistical generalizability. Larger and more diverse samples are needed to test whether coordination patterns differ systematically across older adults and informal caregivers. Additional research should also investigate how practices such as piggybacking and selective disclosure apply in rural communities, higher socioeconomic households, or cross-cultural settings.

We did not probe participants to imagine agent integration. This was intentional to document existing coordination practices without introducing speculative technologies that might bias their accounts. Further research is needed to engage participants directly (e.g., co-design, prototype testing) to understand how they envision, interpret, and negotiate agent involvement.

\section*{Acknowledgement}
We thank the members and staff of the senior centers where we conducted our research and our colleagues across the NSF AI-CARING Institute for their feedback. This research is supported by the National Science Foundation (IIS-2112633).

\bibliographystyle{ACM-Reference-Format}
\bibliography{ref}



\end{document}